\documentclass[aps,
reprint,
prl,
superscriptaddress,
amsfonts,amsmath,amssymb]{revtex4-1}

\usepackage{graphicx}
\usepackage{bm}
\usepackage{amsmath}

\usepackage{hyperref}

\usepackage{pdfpages}
\usepackage{pgffor}

\usepackage{color}
\usepackage{soul}

\DeclareUnicodeCharacter{2212}{-}

\definecolor{OGreen}{rgb}{0,0.6,0}

\usepackage{listings}

\makeatletter
\AtBeginDocument{\let\LS@rot\@undefined}
\makeatother

\begin{document}
\title{Frequency-dependent impedance of nanocapacitors from electrode charge fluctuations as a probe of electrolyte dynamics}
\author{Giovanni Pireddu}
\affiliation{Sorbonne Universit\'{e}, CNRS, Physico-chimie des \'Electrolytes et Nanosyst\`emes Interfaciaux, PHENIX, F-75005 Paris}
\author{Benjamin Rotenberg}
\email{benjamin.rotenberg@sorbonne-universite.fr}
\affiliation{Sorbonne Universit\'{e}, CNRS, Physico-chimie des \'Electrolytes et Nanosyst\`emes Interfaciaux, PHENIX, F-75005 Paris}
\affiliation{R\'eseau sur le Stockage Electrochimique de l'Energie (RS2E), FR CNRS 3459, 80039 Amiens Cedex, France}

%%%%%%%%%%%%%%%%%%%%%%%%%%%%%%%%%%%%%%%%%%%%%%%%%%%%%%%%%%%%%%%%%%%%%%%%%%%%%%%%%%%%%%%%%%%%%

\begin{abstract}
The frequency-dependent impedance is a fundamental property of electrical components. We show that it can be determined from the equilibrium dynamical fluctuations of the electrode charge in constant-potential molecular simulations, extending in particular a fluctuation-dissipation for the capacitance recovered in the low-frequency limit and provide an illustration on water/gold nanocapacitors. This work opens the way to the interpretation of electrochemical impedance measurements in terms of microscopic mechanisms, directly from the dynamics of the electrolyte, or indirectly via equivalent circuit models as in experiments.
\end{abstract}

\maketitle

%%%%%%%%%%%%%%%%%%%%%%%%%%%%%%%%%%%%%%%%%%%%%%%%%%%%%%%%%%%%%%%%%%%%%%%%%%%%%%%%%%%%%%%%%%%%%

The electrical impedance, which quantifies the current induced by a small oscillatory applied voltage, is a fundamental property of the components of an electric circuit. Electrochemical impedance spectroscopy techniques~\cite{mei_physical_2018,wang_electrochemical_2021} are routinely used to characterize batteries~\cite{gaberscek_understanding_2021} and capacitors~\cite{segalini_qualitative_2010}. Electrodes in nanofluidic devices also open new avenues for energy conversion from salinity gradients~\cite{siria2013a,siriaNewAvenuesLargescale2017} or from light~\cite{xiao_artificial_2019}, for catalysis~\cite{levin_nanofluidic_2019} or sensing charged species and measuring ultra-low flow rates with electrochemical correlation spectroscopy~\cite{zevenbergen_electrochemical_2009,mathwig_electrical_2012}. Measurements are usually interpreted in terms of equivalent circuits (EC)~\cite{taberna2003a} combining elements with impedance $Z(\omega)=1/iC\omega$, $R$, or $iL\omega$, with $\omega$ the frequency, for capacitors, resistors or inductors, respectively, or more complex ones such as constant phase elements. The frequency-dependent response highlights the relevant time scales in the dynamics of the polarization of the liquid and the transport of charge carriers within such electrochemical devices. The diffusion and migration of ions can be modeled analytically at the mean-field level~\cite{bazant2004a,barbero_theory_2008,janssen_transient_2018} and their coupling with hydrodynamic flows with mesoscopic simulations methods~\cite{asta2019a}, to predict the charging dynamics of a capacacitor. The effect of steric and other correlations on the dynamics can be introduced in such descriptions~\cite{kilic_steric_2007} or in classical Density Functional Theory~\cite{jiang_kinetic_2014,lian_time-dependent_2016}. Charge transport within porous electrodes, which are used in many devices, can be modeled using such liquid state theories or simpler constitutive equations, in order to make the link with EC such as the transmission line model (TLM)~\cite{biesheuvel_nonlinear_2010,lian_blessing_2020,lin_microscopic_2022}.

Molecular simulations have greatly contributed to tecent progress in the description of electrode/electrolyte interfaces, taking into account the polarization of the electrodes by the ions and solvent molecules of the electrolyte~\cite{scalfi_molecular_2021}. Several methods have been developed to impose the potential of the metal both in \textit{ab initio} and classical molecular dynamics (MD) simulations~\cite{siepmann1995a, reed_electrochemical_2007, bonnet_first-principles_2012,dufils_simulating_2019, deisenbeck_dielectric_2021}, which enabled the molecular-scale study of simple and complex electrochemical systems, including supercapacitors~\cite{merlet2012a, salanne_efficient_2016, simoncelli_blue_2018, jeanmairet_microscopic_2022}. In such constant-potential simulations, the charge of the electrodes fluctuates in response to the dynamical evolution of the electrolyte. The local and global responses of the charge of porous electrodes to an applied voltage step can be monitored to build EC models such as the TLM in order to extrapolate to larger system sizes and longer time scales and bridge the gap with the electrochemical impedance measurements~\cite{pean2014a,pean2016b}, or to design optimal charging protocols~\cite{breitsprecher2018a,breitsprecher_how_2020}. The capacitance and so-called equivalent series resistance of the system can also be obtained using a recently proposed nonequilibrium ``computational amperometry'' approach~\cite{dufils_computational_2021}.

An alternative strategy to compute the capacitance from MD simulations and to extract information on the interfacial electrolyte is to study the equilibrium fluctuations of the electrode charge~\cite{limmer_charge_2013,scalfi2020a}. The differential capacitance, \textit{i.e.} the response of the electrode charge $Q$ to a change in voltage $\Delta\Psi$ between the electrodes, is related to the variance of the charge distribution via a fluctuation-dissipation relation $C_{\rm diff}= \frac{\partial\langle Q \rangle}{\partial \Delta\Psi} = \beta \langle \delta Q^2 \rangle$, where brakets denote an ensemble average, $\delta Q=Q-\langle Q\rangle$ and $\beta=1/k_B T$ with $k_B$ the Boltzmann constant and $T$ the temperature. The link between the thermal fluctuations of charge carriers and electrical response, already observed almost a century ago by Nyquist and Johnson who considered the thermal noise in electronic conductors~\cite{nyquist1928a,johnson1928a}, allowed to correlate using molecular simulations the voltage-dependence of the differential capacitance with changes in the interfacial structure of an ionic liquid or a water-in-salt-electrolyte between graphite electrodes~\cite{merlet2014a,rotenberg2015a,li2018b}. More recently the charge fluctuations within electrodes were also linked to the electrode-electrolyte interfacial free energy~\cite{scalfi_microscopic_2021}.

Here, we show that the frequency-dependent impedance can be extracted from the equilibrium dynamical fluctuations of the electrode charge in constant-potential simulations, the above fluctuation-dissipation relation for the capacitance being recovered in the low-frequency limit. This opens the possibility to investigate with MD simulations the link between the impedance and the dynamics of the liquid between the electrodes. We then illustrate this on water/gold nanocapacitors. Both experiments and simulations recently highlighted some peculiarities of nanoconfined water, such as the decrease of dielectric constant~\cite{fumagalli_anomalously_2018,olivieri_confined_2021} and its anisotropy close to interfaces~\cite{loche_breakdown_2018}, as well as its unusual dynamical behavior~\cite{gekle_anisotropy_2012,mondal_anomalous_2021}. In addition, the dynamics of water on metals proceeds by local reorientation jumps or by rare collective fluctuations, depending on the  metal~\cite{zhang2020a,limmer_hydration_2013,limmer_water_2015}. Finally, the coupling between the interfacial fluctuations of water and the electronic response of carbon plays an important role in the water-graphite friction~\cite{kavokine_fluctuation-induced_2022}. The present approach offers a way to probe the  dynamics of the liquid encoded in the electrical impedance measured with electrodes.

%%%%%%%%%%%%%%%%%%%%%% Figure 1
\begin{figure}[hbt!]
\centering
  \includegraphics[width=3.37in]{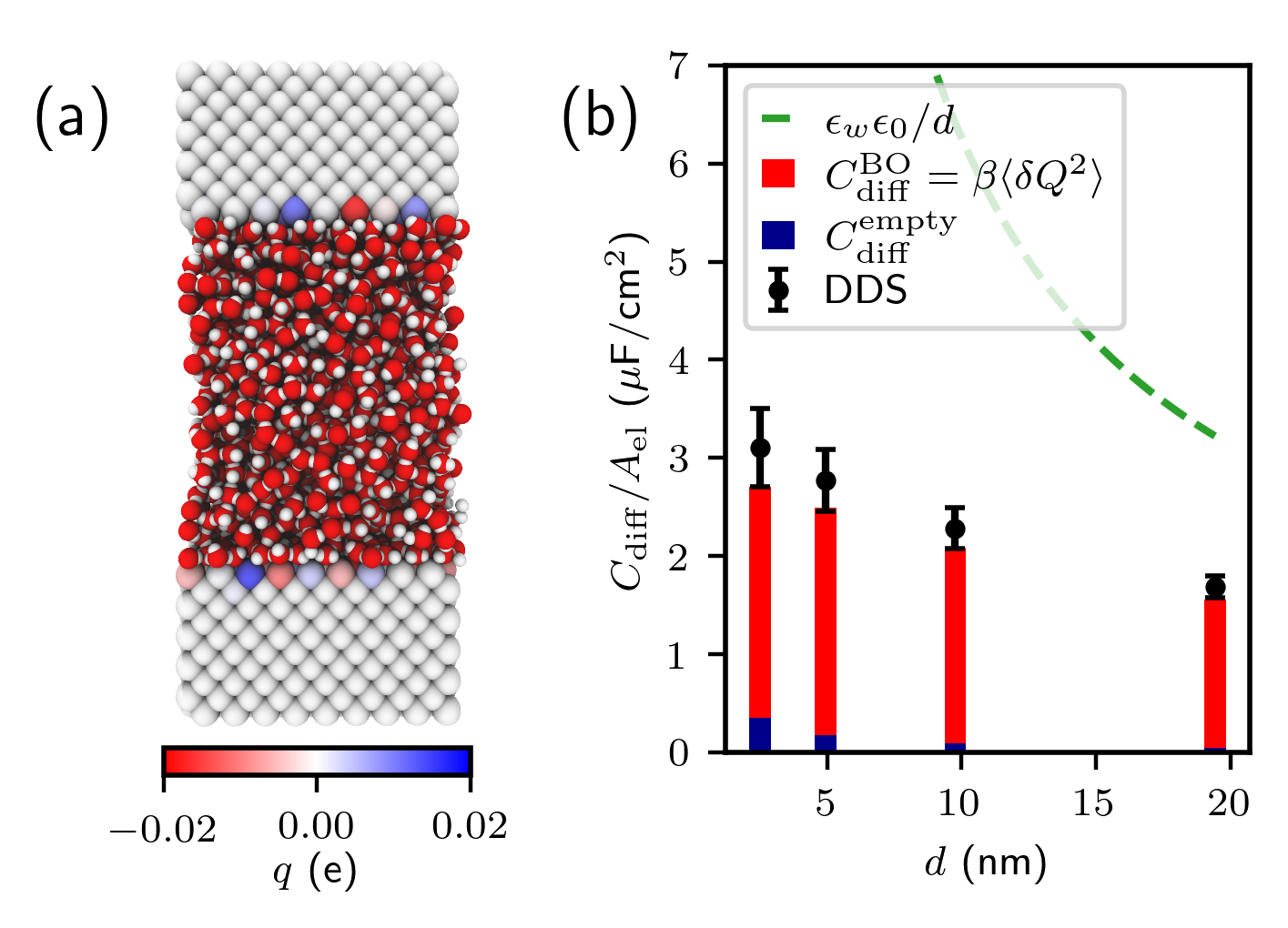}
  \caption{
  (a) Simulated system for an inter-electrode distance $d=4.94$~nm, with the color of gold electrode atoms indicating their instantaneous charge, and water hydrogen and oxygen atoms shown in white and red, respectively.
  (b) Differential capacitance divided by the lateral area of the electrodes, $A_{\rm el}$, as a function of $d$, with the contributions $C_{\rm diff}^{\rm BO}$ arising from the thermal fluctuations of the liquid and $C_{\rm diff}^{\rm empty}$ arising from those suppressed in the Born-Oppenheimer sampling of the charges, which corresponds to the empty capacitor. Also shown are the predictions of continuum electrostatics without (green dashed line) and considering (symbols) the Dielectric Dividing Surfaces (see text; the error bars on the capacitance arise from that on the positions of the DDS).
  }
\label{fig:snap_cdiff}
\end{figure}

We consider systems such as illustrated in Fig.~\ref{fig:snap_cdiff}a, with two rigid electrodes maintained at a fixed potential difference $\Delta\Psi$ and separated by a liquid with fixed number of particles and temperature $T$. The charge of each electrode atom fluctuates in order to satisfy the constant-potential constraint, so that the total charge $\pm Q$ of the oppositely charged electrodes (due to global electroneutrality), which is the thermodynamic variable conjugate to $\Delta\Psi$, also fluctuates in response to the thermal fluctuations of the liquid. The complex admittance $Y(\omega)=1/Z(\omega)$, with $Z(\omega)$ the impedance, quantifies the linear response of the induced electric current, $I(t)=\dot{Q}(t)$, to a small oscillatory voltage, $V(t)= V^0 e^{i\omega t}$, in addition to the voltage $\Delta\Psi$ defining the thermodynamic ensemble, from the change in current $\langle \Delta I (t) \rangle  = Y(\omega) V(t)$. Using linear response theory, we show in Supplemental Material (SM)~\cite{suppmat_pireddu2022} that the admittance can be expressed using the Laplace transform of the current or charge auto-correlation functions (ACF), as:
\begin{align} \label{eq:admittancecharge}
    Y(\omega) &=  \beta \int_{0}^{\infty} \langle \delta I(0) \delta I(t) \rangle e^{-i\omega t}{\rm d}t 
    \nonumber \\
    &=  \beta \left[ i\omega \langle \delta Q^2 \rangle + \omega^2 \int_{0}^{\infty}\langle \delta Q(0) \delta Q(t) \rangle e^{-i\omega t}{\rm d}t \right] .
\end{align}
This fundamental result explicitly links the dynamics of the electrode charge, which reflect the thermal fluctuations of the liquid, with the impedance of the system. It generalizes the above fluctuation-dissipation relation for the capacitance, which is recovered in the low frequency limit $\omega \rightarrow 0$ since the leading term of Eq.~\ref{eq:admittancecharge} is $i \omega \beta \langle \delta Q^2 \rangle$, the admittance of a capacitor with capacitance $\beta \langle \delta Q^2 \rangle$. Importantly, Eq.~\ref{eq:admittancecharge} also offers a practical tool to compute the full impedance spectrum from the dynamics of the charge fluctuations in constant-potential MD simulations, as shown below. 

The fluctuating charges introduced in Ref.~\cite{siepmann1995a} as additional degrees of freedom to simulate constant-potential conditions followed an equation of motion with a fictitious mass. Ref.~\cite{reed_electrochemical_2007} then suggested to determine the set of charges that self-consistently satisfy the constant-potential constraints at each step of the simulation, so that the electrode atom charges are enslaved to the microscopic configurations of the liquid. Such a Born-Oppenheimer (BO) sampling allows to use larger time steps, but also suppresses some charge fluctuations that may contribute to some observables. For example, the suppressed fluctuations lead to a missing term in the capacitance corresponding to that of an empty capacitor, $C_{\rm diff}^{\rm empty}$~\cite{scalfi2020a}. Separating the instantaneous charge as $Q(t)=Q^{\rm BO}(t)+Q^{\rm nBO}(t)$ into the BO and non-BO terms and expanding $\langle \delta Q^2 \rangle$ and  $\langle \delta Q(0) \delta Q(t) \rangle$ in Eq.~\ref{eq:admittancecharge} yields
\begin{align}\label{eq:admittancecharge2}
    Y(\omega)= Y^{\rm BO}(\omega) + Y^{\rm nBO}(\omega) \ ,
\end{align}
where the first term is obtained by replacing $Q$ by $Q^{\rm BO}$ in Eq.~\ref{eq:admittancecharge} and the second is the remainder. The latter depends on the choice of dynamics for the additional degrees of freedom, which should only manifest themselves at high frequencies since they mimick the electronic response, except for the ``static'' contribution for $\omega \rightarrow 0$  where $Y^{\rm nBO}(\omega)\approx i\omega C_{\rm diff}^{\rm empty}$. In practice, the non-BO contribution to the capacitance is small compared to the BO one arising from the thermal fluctuations of the liquid~\cite{scalfi2020a} (see also below). In the following, we drop the BO superscript and refer to the BO charges as simply $Q$. 

As an illustration of the possibilities offered by the analysis of the equilibrium dynamical fluctuations of the electrode charge in constant-potential simulations, we investigate gold/water nanocapacitors (see Fig.~\ref{fig:snap_cdiff}a). Simulations are performed under periodic boundary conditions in the $x$ and $y$ directions only, with fixed number of molecules and corresponding inter-electrode distance. The latter is determined by prior equilibration at a constant pressure of 1~atm in the direction perpendicular to the electrode surfaces, resulting in $d=$2.51, 4.94, 9.76, and 19.42~nm between the first atomic planes of the electrodes. Details on the systems and molecular models can be found in the SM~\cite{suppmat_pireddu2022}. All simulations are performed using the MD code Metalwalls~\cite{marin-lafleche_metalwalls_2020}, with a potential difference $\Delta\Psi=0$~V between the electrodes except for the response to a voltage step. Unless otherwise stated, the uncertainty is expressed as standard error computed from 10 blocks of the trajectory.

Fig.~\ref{fig:snap_cdiff}b shows the BO and non-BO contributions to the differential capacitance as a function of the inter-electrode distance $d$. As indicated above, the former contribution, $C_{\rm diff}^{\rm BO}=\beta \langle \delta Q^2 \rangle$, is much larger than the latter, $C_{\rm diff}^{\rm empty}$, computed from the properties of the electrodes as described in Ref.~\citenum{scalfi2020a}. The figure also shows that the prediction from continuum electrostatics, $\varepsilon_w\varepsilon_0 A_{\rm el}/d$, with $\varepsilon_0$ the vacuum permittivity, $\varepsilon_w$ the relative permittivity of bulk water, and $A_{\rm el}$ the lateral area of the electrodes, only qualitatively reproduces the decrease of the capacitance with increasing inter-electrode distance. This is due to the failure to capture the potential drop across the interfacial water layers, as previously reported on other surfaces~\cite{jeanmairet2019b}. The prediction can be greatly improved by considering the Dielectric Dividing Surface (DDS), obtained from the interfacial permittivity profiles~\cite{schlaich_water_2016}. As the Gibbs Dividing Surface for density profiles, the DDS locates an equivalent sharp interface between a (water-free) region with permittivity $\varepsilon_0$ and another with $\varepsilon_0\varepsilon_w$ (see SM~\cite{suppmat_pireddu2022}). In this picture, the capacitance per unit area corresponds to that of three capacitors in series, $A_{\rm el}/C_{\rm DDS}= 2 w_{\rm DDS} / \varepsilon_0 + d_{\rm DDS} / \varepsilon_0\varepsilon_w$, with $w_{\rm DDS}\approx 1.27$~\AA\ the distance between the last atomic plane of each electrode and the corresponding DDS, and $d_{\rm DDS}= d-2 w_{\rm DDS}$ the width of the equivalent bulk water region, and provides a rather accurate description of the actual capacitance (see Ref.~\citenum{cox_dielectric_2022} for a related approach for the solvation free energy). 

%%%%%%%%%%%%%%%%%%%%%% Figure 2

\begin{figure}[hbt!]
\centering
  \includegraphics[width=3.37in]{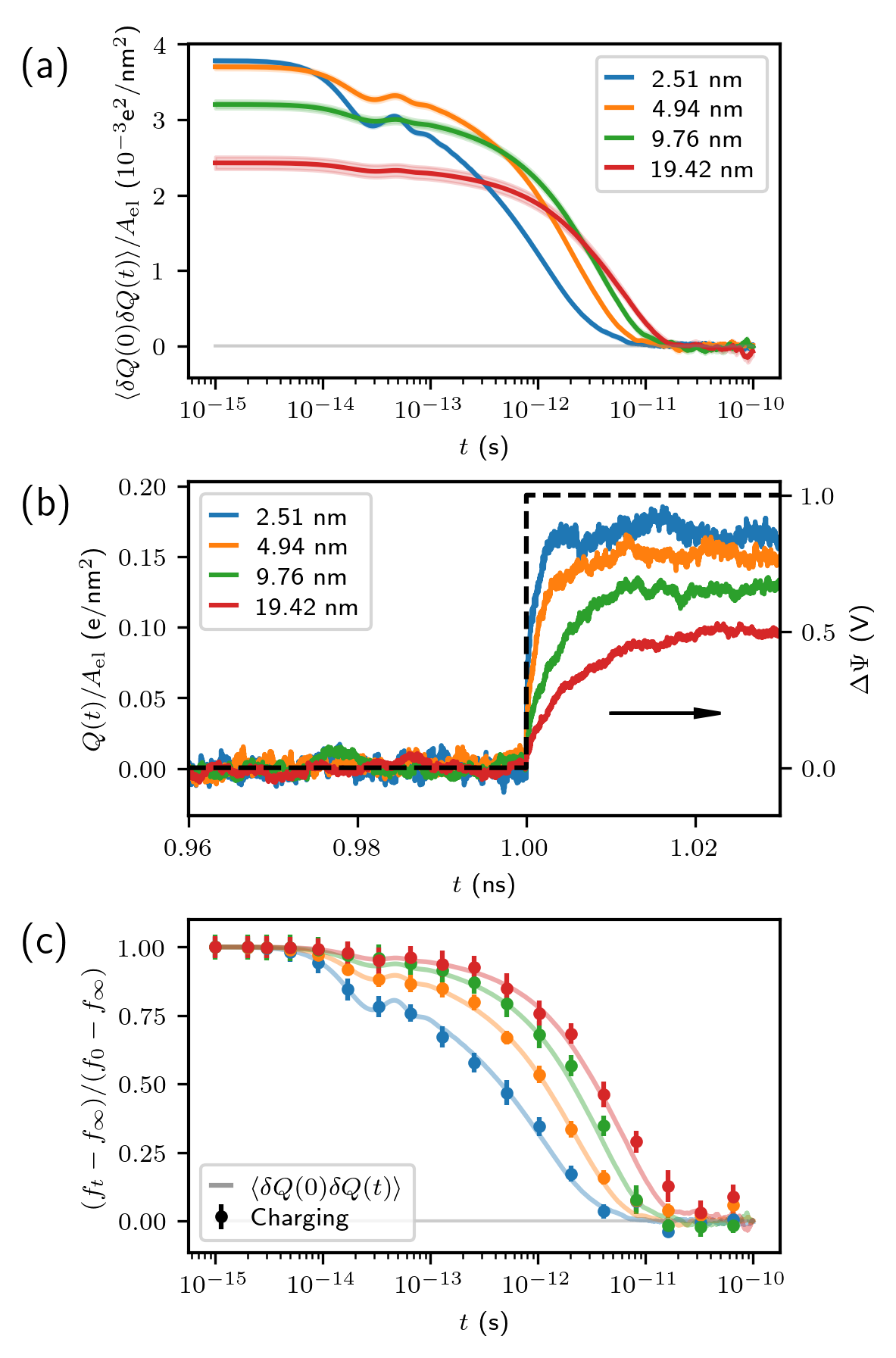}
  \caption{
  (a) Electrode charge auto-correlation function, $\langle \delta Q(0) \delta Q(t) \rangle$, divided by $A_{\rm el}$ for the 4 considered inter-electrode distances; the shaded areas represent the standard error associated with each curve.
  (b) Response of the charge $Q$ divided by $A_{\rm el}$ (colored solid lines) to a voltage change from $\Delta\Psi=0$~V to 1~V (black dashed line). For each system, the results are averaged over 10 trajectories.
  (c) Comparison of the responses with the equilibrium charge ACFs, normalized as $(f_t-f_\infty)/(f_t-f_\infty)$, where $f$ is either $Q$ recorded during the charging (symbols), or $\langle \delta Q(0) \delta Q(t) \rangle$ (solid lines).
  }
\label{fig:charging}
\end{figure}

Turning to the dynamics of the charge fluctuations, Fig.~\ref{fig:charging}a shows that the charge ACF displays oscillations at short times and an approximately exponential decay at longer times. Increasing the inter-electrode distance damps the former and slows down the latter. In the linear response regime, $\langle \delta Q(0) \delta Q(t) \rangle$ should reflect the evolution of $Q$ during the (dis)charge of the capacitor. Fig.~\ref{fig:charging}b shows the charge as a function of time before after a change from $\Delta\Psi=0$~V to 1~V. The plateaus reached after the transient regime are consistent with the capacitance for each $d$, and the characteristic time to reach the new equilibrium increases with $d$, consistently with the slower decay of the ACF in Fig.~\ref{fig:charging}a. However, the link between the equilibrium ACF and the out-of-equilibrium response is much deeper: Fig.~\ref{fig:charging}c shows the equivalence between the two when properly rescaled, as expected from the fluctuation-dissipation theorem, indicating that in these systems the response is linear at least up to 1~V (as observed for the capacitance of water/graphite capacitors~\cite{jeanmairet2019b}).

%%%%%%%%%%%%%%%%%%%%%% Figure 3

\begin{figure}[hbt!]
\centering
  \includegraphics[width=3.37in]{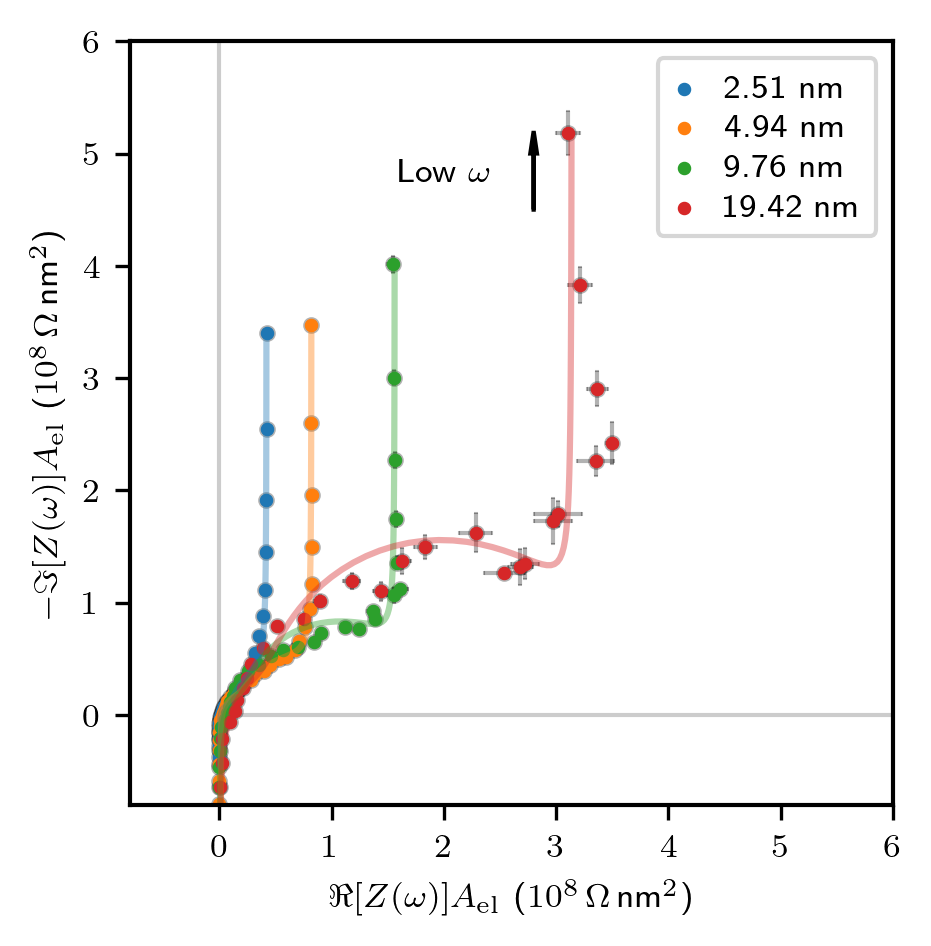}
  \caption{Nyquist plot of the impedance $Z(\omega)$ in the complex plane, parameterized by the frequency $\omega$, for the 4 considered inter-electrode distances (colors). Results from MD simulations using Eq.~\ref{eq:admittancecharge} (symbols, with error bars indicating the standard error) are shown with the predictions of the corresponding equivalent circuit models (solid lines, see text and SM~\cite{suppmat_pireddu2022}).
  }
\label{fig:Nyquist}
\end{figure}

The impedance is usually analyzed in a Nyquist plot, which reports its imaginary and real parts in the complex plane, parameterized by the frequency. The MD results obtained from the ACF using Eq.~\ref{eq:admittancecharge} (see SM~\cite{suppmat_pireddu2022} for more details on the numerical procedure) are shown in Fig.~\ref{fig:Nyquist}, for a frequency range spanning 4 orders of magnitude, from $1.25\times 10^{11}$ to $1.8\times 10^{15}$~rad~s$^{-1}$. For a sufficiently large system in the lateral directions, the charge ACF, hence the admittance scales linearly with $A_{\rm el}$, as shown numerically in SM~\cite{suppmat_pireddu2022}. We therefore plot the impedance multiplied by $A_{\rm el}$ in Fig.~\ref{fig:Nyquist}.
As in experiments, we model the data using an EC with resistances and capacitances fitted to the admittance/impedance from MD (see SM~\cite{suppmat_pireddu2022}). The Nyquist plots are typical of capacitors with a vertical asymptote at low frequency. The growing real part with inter-electrode distance reflects an increase in dissipative processes. The corresponding resistance scales approximately as $d/A_{\rm el}$,  as expected for a macroscopic resistor. At the other end of the frequency spectrum, we observe a sign change in the imaginary part, which reflects the short-time oscillations of the ACF and can be modeled using an inductor in the EC, with $L\propto d/A_{\rm el}$.

%%%%%%%%%%%%%%%%%%%%%% Figure 4

\begin{figure}[hbt!]
\centering
  \includegraphics[width=3.37in]{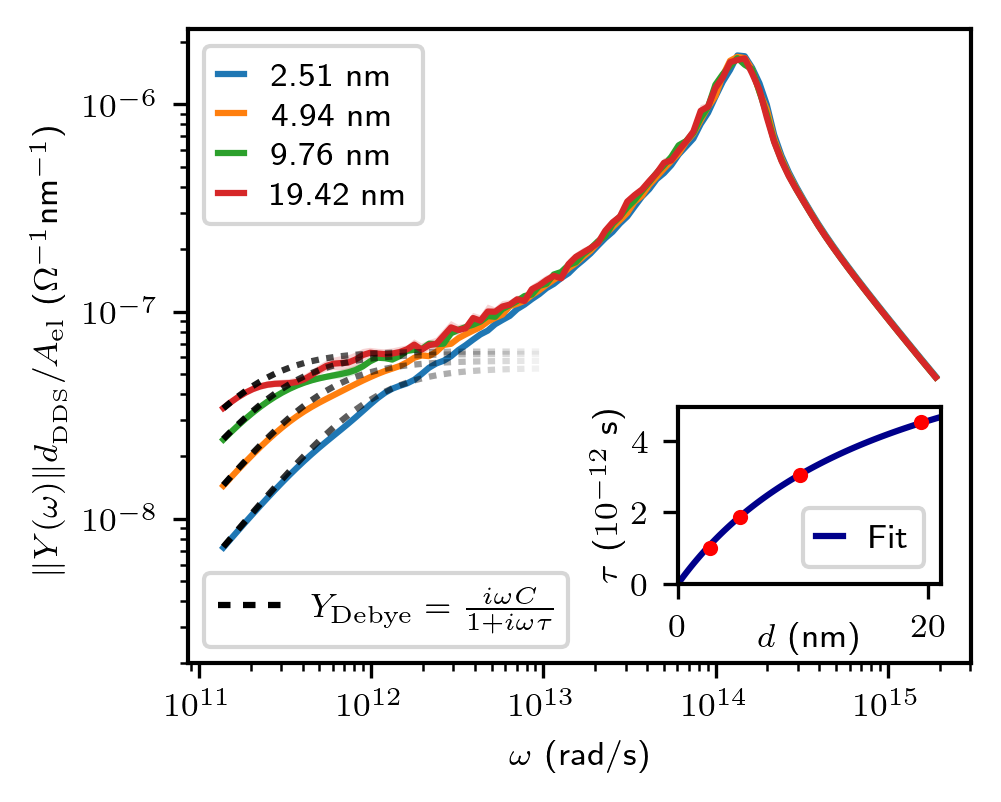}
  \caption{
  Norm of the complex admittance obtained via Eq.~\ref{eq:admittancecharge} for the 4 considered inter-electrode distances (colors). The MD simulation results (solid lines) are shown multiplied by the distance $d_{\rm DDS}$ between the two Dielectric Dividing Surfaces and divided by the lateral area $A_{\rm el}$ to highlight the scalings with system size. The shaded area represent the standard error. The dashed lines indicate the prediction at low frequency of a mono-exponential model using the capacitance $C$ and the characteristic time $\tau$ defined as the integral of the normalized charge ACF. The inset shows  $\tau$ as a function of $d$, with a fit corresponding to the DDS picture (see text).}
\label{fig:Bode}
\end{figure}

The frequency-dependence is shown in the Bode plot, Fig.~\ref{fig:Bode}, where the norm of the complex admittance is multiplied by the effective water slab width $d_{\rm DDS}$ and divided by the lateral area $A_{\rm el}$ to highlight the scaling relations with system size. The most prominent feature is a common peak, with magnitude inversely proportional to $d_{\rm DDS}$, located at $\omega\approx 1.5\times10^{14}$~rad~s$^{-1}$ and corresponding to the short-time oscillations observed in the charge ACF shown in Fig.~\ref{fig:charging}a. Interestingly, this frequency coincides with the one of the previously reported oscillations of the dipole moment of nanoconfined water~\cite{gekle_anisotropy_2012,mondal_anomalous_2021} as well as a feature in the bulk permittivity spectrum of water ascribed to librational modes~\cite{carlson_exploring_2020}. This points to the link between the water polarization and the electrode charge (see below). In addition, the linear scaling of this peak with $d_{\rm DDS}$ suggests that it is mainly due to ``bulk'' water rather than to an interfacial mechanism.

The broad shoulder at low frequency is well described by a Debye model with a single relaxation time. Indeed, from Eq.~\ref{eq:admittancecharge} the admittance can be approximated at low frequency by $Y(\omega)\approx  i \omega \beta \langle \delta Q^2 \rangle \times (1 - i\omega \tau) \approx i \omega C_{\rm diff} / ( 1 + i\omega \tau)$, with $\tau$ the integral of the normalized charge ACF. The deviations from the scaling with $d_{\rm DDS}$ reflect the interfacial component of the capacitance captured by the DDS model (see the above discussion of Fig.~\ref{fig:snap_cdiff}). In the EC picture, this low-frequency behavior corresponds to a RC circuit in series, with relaxation time $\tau=RC$. Both estimates of $\tau$ coincide and increase with inter-electrode distance approximately as $\tau_\infty\times d/[d + 2(\varepsilon_w-1) w_{\rm DDS}]$ (see inset of Fig.~\ref{fig:snap_cdiff} and SM~\cite{suppmat_pireddu2022}). The fitted value $\tau_\infty\approx 8.6$~ps is close to the characteristic time corresponding to the Debye relaxation in bulk water ($\tau_{\rm D}\approx 9$~ps) that can be interpreted as arising from the migration of orientational defects in the H-bond network~\cite{popov_mechanism_2016}, even though the separation between individual and collective motion is challenging~\cite{carlson_exploring_2020}. This further illustrates the link between the impedance and the dynamics of the charge fluctuations in MD simulations. 

As a first step towards the molecular interpretation of the admittance/impedance, we monitor the components of the dipole moment of the confined water slab in the directions perpendicular and parallel to the electrode surfaces. We show in SM~\cite{suppmat_pireddu2022} that the former almost perfectly follows the dynamics of the electrode charge, while the latter decays more slowly and is only slightly affected by $d$. We also show that, under the considered conditions, the perpendicular component of the total water dipole satisfies $M^{\perp}_{\rm wat}(t) = Q(t) d_Q(t)$, with $d_Q$ the instantaneous distance between the charge-weighted atomic positions in each electrode. In the present case of a perfect metal, these positions almost coincide with the first atomic planes for all microscopic configurations (see also Refs.~\citenum{pireddu_molecular_2021,serva_effect_2021} for the charge induced on the same model of gold electrodes), so that $M^{\perp}_{\rm wat}(t) \approx Q(t) d$, which explains the direct link between the ACF of the electrode charge and that of $M^{\perp}_{\rm wat}$. The situation might be different in the case of imperfect screening of charge and potential within metals, which can be introduced in MD using the Thomas-Fermi model~\cite{scalfi_semiclassical_2020}.

%%%%%%%%%% Conclusion

Overall, the present work offers the possibility to extract the frequency-dependent impedance/admittance of electrochemical systems in molecular simulations from the equilibrium dynamical fluctuations of the electrode charge. This opens the way to the interpretation of impedance measurements in terms of microscopic mechanisms, either directly from the dynamics of the electrolyte, or indirectly by EC models used to analyze the experimental data. Beyond the case of pure water considered as an illustration, the next step is to investigate electrolyte solutions and the dynamics of electric double layers. The diffusion of ions introduces longer timescales, so that the study of the charge fluctuations may benefit from implicit solvent descriptions of electrochemical interfaces~\cite{cats_capacitance_2021} and would also bridge the gap with the experimental frequency range probed in typical electrochemical impedance measurements. Nevertheless, the one considered here for pure water already partially overlaps with that of THz spectroscopy experiments, which can also be decomposed into microscopic contributions~\cite{heyden_dissecting_2010} and have recently been used to investigate electrode/electrolyte interfaces~\cite{alfarano_stripping_2021}.  Future directions may also include the non-linear dynamical response to large voltages, using \textit{e.g.} the statistical tools developed to sample electric current fluctuations in MD simulations~\cite{lesnicki_field-dependent_2020,lesnicki_molecular_2021}, as well as systems with redox reactions at the surface of the electrodes~\cite{dwelle2019a}.

%%%%%%%%%% Acknowledgments

The authors thank Mathieu Salanne, Iurii Chubak, Minh-Th\'e Hoang Ngoc and Jeongmin Kim for discussions. This project received funding from the European Research Council under the European Union’s Horizon 2020 research and innovation program (Grant Agreement No. 863473). The authors acknowledge access to HPC resources from GENCI (grant no.~A0110912966). 

%\bibliography{references}


%%%%%%%%%% SM

\clearpage
\ \vfill
\vfill
\foreach \x in {1,2,3,4,5,6,7,8,9,10,11,12,13,14,15,16,17,18,19,20,21,22,23}
{%
\clearpage


\section*{Supplementary material}

# Frequency-dependent impedance of nanocapacitors from electrode charge fluctuations as a probe of electrolyte dynamics
# Supplemental Material

Giovanni Pireddu[1] and Benjamin Rotenberg[1,2,*]

[1]*Sorbonne Université, CNRS, Physico-chimie des Électrolytes et Nanosystèmes Interfaciaux, PHENIX, F-75005 Paris*

[2]*Réseau sur le Stockage Electrochimique de l'Energie (RS2E), FR CNRS 3459, 80039 Amiens Cedex, France*

* benjamin.rotenberg@sorbonne-universite.fr

**CONTENTS**

# I. ADMITTANCE FROM THE ELECTRODE CHARGE FLUCTUATIONS

In this section, we report a more detailed derivation of Eq. 1 in the main text which relates the fluctuations of electrode charges and the admittance. The following derivation is inspired by the one reported in Ref. 1 (see in particular section 7.6), and considers a perturbation of a reference Hamiltonian $\mathcal{H}^0$ written in the form

$$\mathcal{H}_{tot} = \mathcal{H}^0 - AF(t), \tag{1}$$

where $F(t)$ represents the time-dependent control variable perturbing the system and $A$ is its conjugate variable. The response to a generic time-dependent perturbation can be decomposed into its Fourier components, so that it is sufficient to consider the effect of a monochromatic perturbation $F(t) = F^0 e^{i\omega t}$. As described in Ref. 1, one should in principle introduce a regularization in the form of a factor $e^{\epsilon t}$ with $\epsilon > 0$ ensuring the absence of perturbation for $t \to -\infty$ and take the limit $\epsilon \to 0$ at the end of the calculation, but we drop it in the following for simplicity. Linear response theory then allows expressing the average response of a variable $B$ to the (sufficiently small) perturbation as

$$\langle \Delta B(t) \rangle = \int_{-\infty}^{t} \phi_{BA}(t-s) F(s) \mathrm{d}s, \tag{2}$$

where $\phi_{AB}(t)$ is the *after-effect* function, which can be express as the equilibrium correlation function:

$$\phi_{BA}(t) = \beta \langle \delta B(t) \delta \dot{A} \rangle, \tag{3}$$

with $\delta B = B - \langle B \rangle$ (and similarly for $A$) and $\dot{A}$ denotes the time-derivative or $A$.

The frequency-dependent electrical impedance $Z(\omega)$ is the inverse of the admittance $Y(\omega)$, which quantifies the response of the electric current $I = \dot{Q}$ to a change $V$ in the voltage $\Delta \Psi$ between the two oppositely charged electrodes. The conjugate variable of voltage is the total charge $\pm Q$ of the latter. We therefore consider the above general results in the particular case $A = Q$, $B = I$, and $F = V$ (with the appropriate sign convention to define the voltage). Introducing the monochromatic form of $V(t) = V^0 e^{i\omega t}$ in Eq. 2, we obtain:

$$\langle \Delta I(t) \rangle = \int_{-\infty}^{t} \phi_{IQ}(t-s) V(s) \mathrm{d}s = V^0 \int_{-\infty}^{t} \phi_{IQ}(t-s) \exp[i\omega s] \mathrm{d}s$$

$$= V^0 \exp[i\omega t] \int_{-\infty}^{t} \phi_{IQ}(t-s) \exp[-i\omega(t-s)] \mathrm{d}s$$

$$= V^0 \exp[i\omega t] \int_{0}^{\infty} \phi_{IQ}(t) \exp[-i\omega t] \mathrm{d}t = V(t) \int_{0}^{\infty} \phi_{IQ}(t) \exp[-i\omega t] \mathrm{d}t.$$

One can therefore identify the Fourier-Laplace transform of the after-effect function $\phi_{IQ}(t)$ with the electrical admittance defined by $\langle \Delta I(t) \rangle = Y(\omega)V(t)$. From Eq. 3, we obtain the admittance from the autocorrelation function of the current as:

$$Y(\omega) = \beta \int_0^\infty \langle \delta I(0) \delta I(t) \rangle e^{-i\omega t} dt . \qquad (4)$$

An alternative expression as function of the electrode charges can be obtained by noting that $\langle \delta I(0) \delta I(t) \rangle = -d^2 \langle \delta Q(0) \delta Q(t) \rangle / dt^2$, which follows from Faraday's law. Considering Laplace transform properties for derivatives, we obtain

$$Y(\omega) = \beta \left[ i\omega \langle \delta Q^2 \rangle + \omega^2 \int_0^\infty \langle \delta Q(0) \delta Q(t) \rangle e^{-i\omega t} dt \right], \qquad (5)$$

where we omitted a first derivative contribution $d\langle \delta Q(0) \delta Q(0^-) \rangle / dt$. This contribution vanishes since equilibrium autocorrelation functions are even with respect to time, and if we assume that $\langle \delta Q(0) \delta Q(t) \rangle$ is differentiable at $t = 0$. Eq. 5 allows to determine the frequency-dependent electrical impedance of the system from the dynamics of the equilibrium fluctuations of the electrode charge, which can be sampled in molecular dynamics simulations.

## II. SIMULATION DETAILS

### A. Setup

Each system includes two confining electrodes, separated by different distances $d = 2.51$, 4.94, 9.76 and 19.42 nm for 1080, 2160, 4320, and 8640 water molecules, respectively. Each electrode consists of 1620 atoms fixed on an FCC lattice ($9 \times 9 \times 5$ unit cells with a lattice parameter of a $= 4.07$ Å corresponding to the unit cell of gold), resulting in ten atomic planes perpendicular to the $z$ direction and facing the inner part of the system with a (100) plane. For all the systems, the box has lateral dimensions $L_x = L_y = 36.63$ Å, and periodic boundary conditions are applied only in the $x$ and $y$ directions.

All the atoms in the systems interact via electrostatic interactions, computed using a 2D Ewald sum method taking into account the Gaussian distributions of the electrode atoms [2, 3] and truncated and shifted Lennard-Jones (LJ) potentials. Water molecules are modeled with the SPC/E force field [4] and the LJ parameters for the gold atoms are taken from Ref. 5, with the Lorentz–Berthelot mixing rules. Both the electrodes were treated

as metals using the fluctuating charge model, in which each electrode atom is equipped with a Gaussian charge of width $w = 0.40$ Å, with magnitude being determined for each configuration, using the matrix inversion method [6], to fulfill the constraints of constant potential and of global electroneutrality.

Each simulation box is prepared with a preliminary equilibration in the $NP_zT$ ensemble, fixing the constant pressure at 1 atm, to determine the equilibrium electrode-electrode distance. At this stage the electrodes act as pistons, and the system box is allowed to elongate or compress along the $z$ direction only. After equilibrating the electrode positions, we run a short $NVT$ simulation to allow the system to fully equilibrate in the conditions used for production. In total, each equilibration includes at least 1 ns of dynamics. In all the simulations, we use a 1 fs time step and the temperature is fixed at 298 K using a Nosé–Hoover chain thermostat [7] with a time constant of 1 ps. All simulations are performed using the molecular dynamics code Metalwalls [8].

### B. Equilibrium and nonequilibrium simulations

The simulations at equilibrium conditions with a voltage $\Delta\psi = 0$ V are conducted for at least 10 ns for production. During the simulations, we sample both the total electrode charge $Q$ and the components of the total dipole moment at each time step. Molecular configurations are sampled each picosecond. From these equilibrium trajectories, we also extract ten configurations separated by 1 ns which are then used as initial configurations for the non-equilibrium simulations. The nonequilibrium response is obtained, for each of these initial configurations, by performing simulations under a voltage $\Delta\psi = 1$ V for at least 250 ps, sampling the total charge at every time step.

### C. Effect of lateral dimensions

In order to check the effect of the lateral dimensions of the simulation box $L_x$ and $L_y$, we perform a supplemental simulation doubling the dimensions in the two directions, for the system with an electrode-electrode separation $d = 4.94$ nm. The electrode atoms are replicated in the two directions and the number of water is four times the original amount. Fig. S1 compares the total charge autocorrelation function for the original and larger systems,

scaled by the lateral area. The good agreement between the two shows that the original system is already sufficiently large.

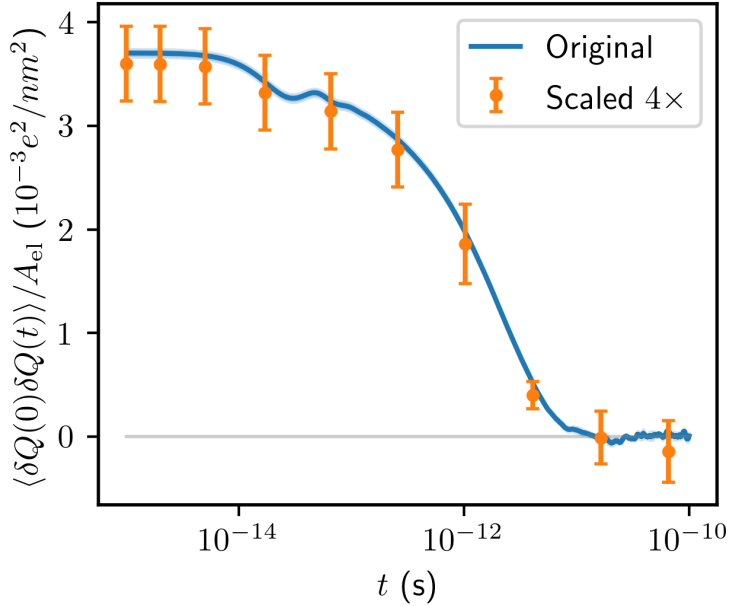

FIG. S1. Total charge autocorrelation functions for the original system (solid lines), and its scaled version (dots), obtained by doubling $L_x$ and $L_y$. The results have been scaled by the respectives electrode areas $A_{el}$ to facilitate the comparison. The uncertainty is calculated as the standard deviation, obtained splitting the original data in 10 segments.

## III.  STATIC PROPERTIES

### A.  Total charge distributions

The total charge histograms are shown in Fig. S2 for all the systems we considered, at 0 V. All the histograms present a Gaussian shape, with the variance being inversely proportional to the electrode-electrode distance.

### B.  Equilibrium density profiles

The equilibrium water density profiles along the $z$-direction are reported in Fig S3. All the systems feature the same layered water structure close to the electrode surface. In

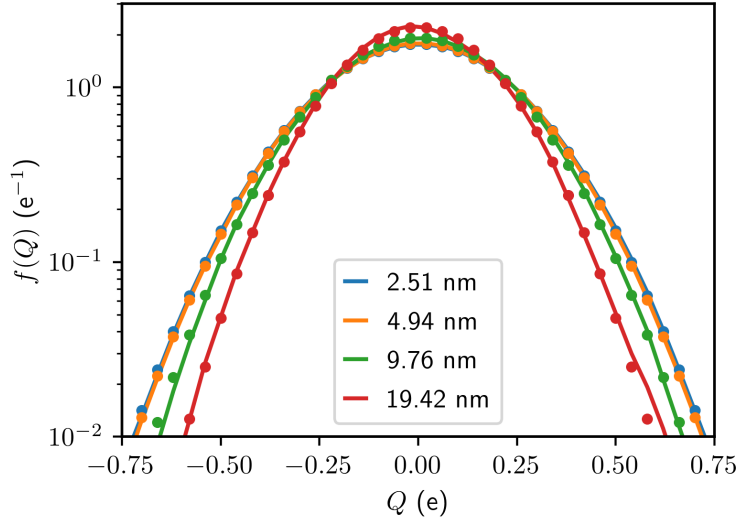

FIG. S2. Distribution of total electrode charge for all the systems we considered, at 0 V. The lines are the results from MD simulations, the dots represent Gaussian probability density functions with zero mean and the variance set to be $\langle Q^2 \rangle$ as calculated from MD results.

general, the density profiles differ only in terms of width of the plateau region, which is trivially wider for larger systems, except for the smallest system we considered. In the latter case, the density profile does not reach a plateau, but the layered structure extends through the whole width of the system.

### C. Dielectric profile

In order to characterize the local dielectric properties of our systems, we computed the dielectric profiles $\varepsilon_\perp^{-1}(z)$ in the confining direction (perpendicularly to the electrode planes). We estimated the profiles using the approach described by Schlaich et al. described in Ref. 9. The results of our analysis are reported in Fig. S4.

All the systems show the same identical charge density profile, shown in Fig. S4a. Based on this observation, we consider the dielectric profile only for the case with $d = 4.94$ nm without loss of generality. The dielectric profile for interfacial water shows an oscillating behaviour close to the gold surface, as expected. From the dielectric profile, we calculated

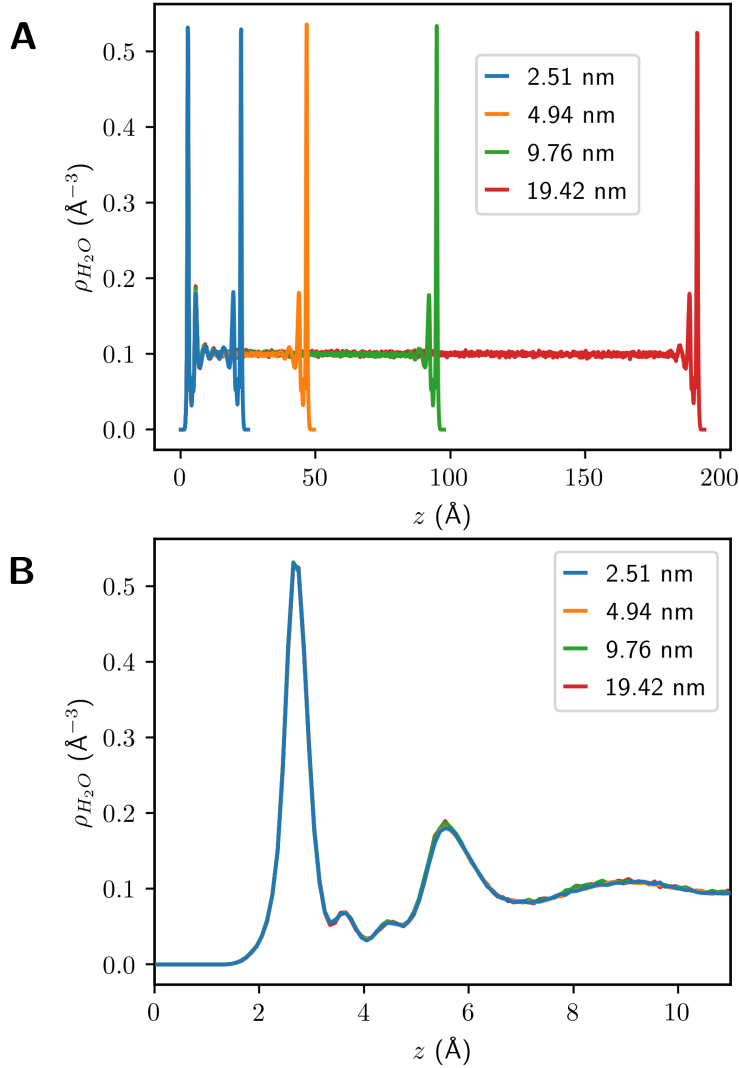

FIG. S3. Water density profiles along the $z$-direction. (a) Comparison between the different systems. (b) Zoom on the leftmost part of the profiles.

the position of the Dielectric Dividing Surface (DDS) using the following definition [9]:

$$z^{\text{DDS}} = z_v + \int_{z_v}^{z_l} \frac{\varepsilon_\perp^{-1}(z_l) - \varepsilon_\perp^{-1}(z)}{\varepsilon_\perp^{-1}(z_l) - \varepsilon_\perp^{-1}(z_v)} dz \qquad (6)$$

where $z_v$ is the position of the electrode surface ($z = 0$ in our reference frame), $z_l$ is a position in the water phase. In practice, we calculated $\varepsilon_\perp^{-1}(z_l)$ as the average value of $\varepsilon_\perp^{-1}(z)$ in the profile plateau. In our case, we use $z^{\text{DDS}}$ to determine the effective water slab width $d_{\text{DDS}} = d - 2w_{\text{DDS}}$, where $w_{\text{DDS}} = z^{\text{DDS}} - z_v \approx 1.27$ Å. The water slab width is used to highlight the scaling relations between the admittance values in systems with varying $d$. We also define a coarse-grained version of the dielectric profile using a step function based on

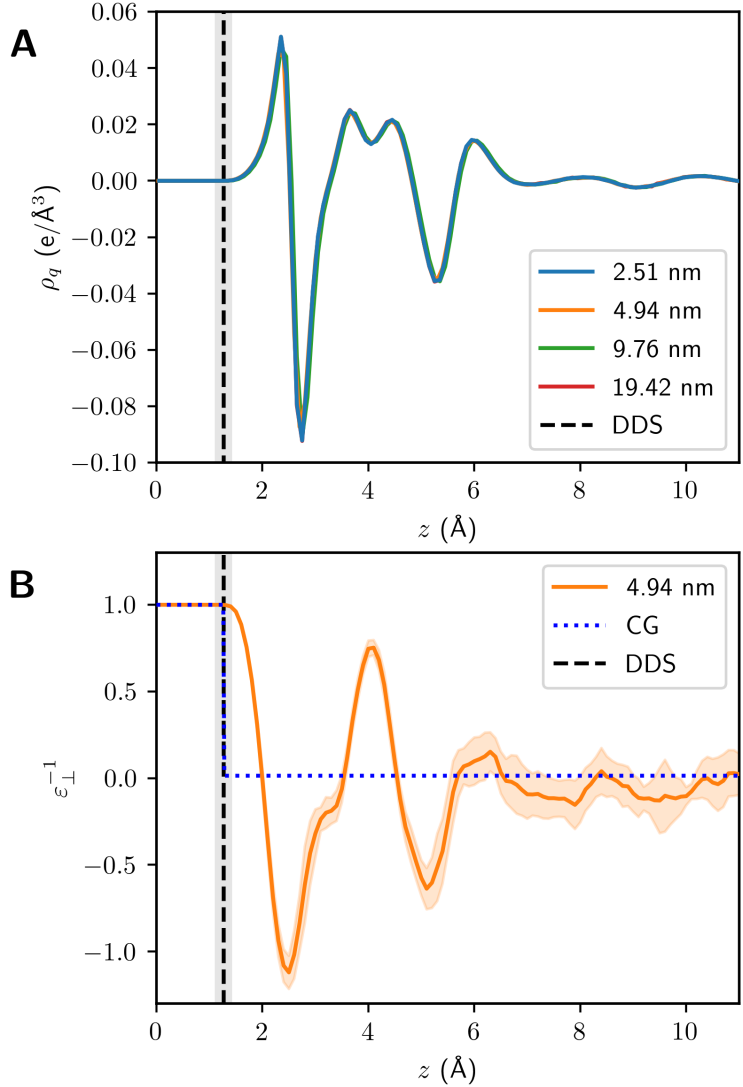

FIG. S4. Charge density and dielectric profiles along the $z$-direction. (a) average charge density profiles for all the systems we considered. (b) dielectric profile for the system with $d = 4.94$. The dashed vertical line indicates the position of the dielectric dividing surface (DDS). The dotted line represents the coarse-grained version of the dielectric profile, defined as a step function and based on the DDS position. The shaded areas indicate the uncertainties, computed as the standard deviation among 10 different samples drawn from the original trajectory.

the position of the DDS [10]

$$\varepsilon_\perp(z) = \begin{cases} 1, & \text{for } z \leq z^{\text{DDS}} \\ \varepsilon_w, & \text{for } z > z^{\text{DDS}} \end{cases}, \tag{7}$$

where $\varepsilon_w$ is the dielectric constant of bulk water, which corresponds to 70.7 in the specific case of SPC/E water. This simple model is used to rationalize the capacitance results presented in the main text.

## IV. CALCULATION OF THE IMPEDANCE FROM MD SIMULATIONS

### A. Windowing

Time autocorrelation functions estimated from MD results suffer from the presence of noise at large $t$ values, because of the limited sampling. This effect impacts negatively on the numerical calculation of the Laplace transform for the estimation of admittance/impedance. In particular, we found this noise to particularly deteriorate the results at high frequency. To suppress this effect, we treated the charge autocorrelation functions by multiplying them with sigmoidal *window* functions having the following form

$$W(t, \epsilon, \tau) = \frac{1}{1 + e^{\epsilon(t-\tau)}}, \quad (8)$$

with $\epsilon$ and $\tau$ two tunable parameters which were adjusted by hand in order to remove the long-time noise while minimizing the impact on the overall shape of the autocorrelation functions. It is clear that this kind of restriction of the original signal suppresses all the slow modes whose period is longer than the region selected by the window function. However, we tuned the window parameters such that only the most significant part of the autocorrelation function was taken into account. A comparison between the original and windowed versions of the autocorrelation functions is reported in Fig. S5 for all the systems. Once the parameters have been adjusted for each system, the same windowing functions were used to treat every data segment used for the estimation of $\langle \delta Q(0) \delta Q(t) \rangle$ and admittance/impedance.

### B. Filon-Lagrange integration

The estimation of admittance/impedance involves the numerical calculation of the Fourier-Laplace transform of the electrode charge autocorrelation function $\langle \delta Q(0) \delta Q(t) \rangle$. For the numerical calculation of the Laplace transforms, and of highly oscillatory integrals in general, general-purpose quadrature methods such as the trapezium or Simpson's rule, may fail in terms of numerical stability, especially at high frequencies. In principle, one

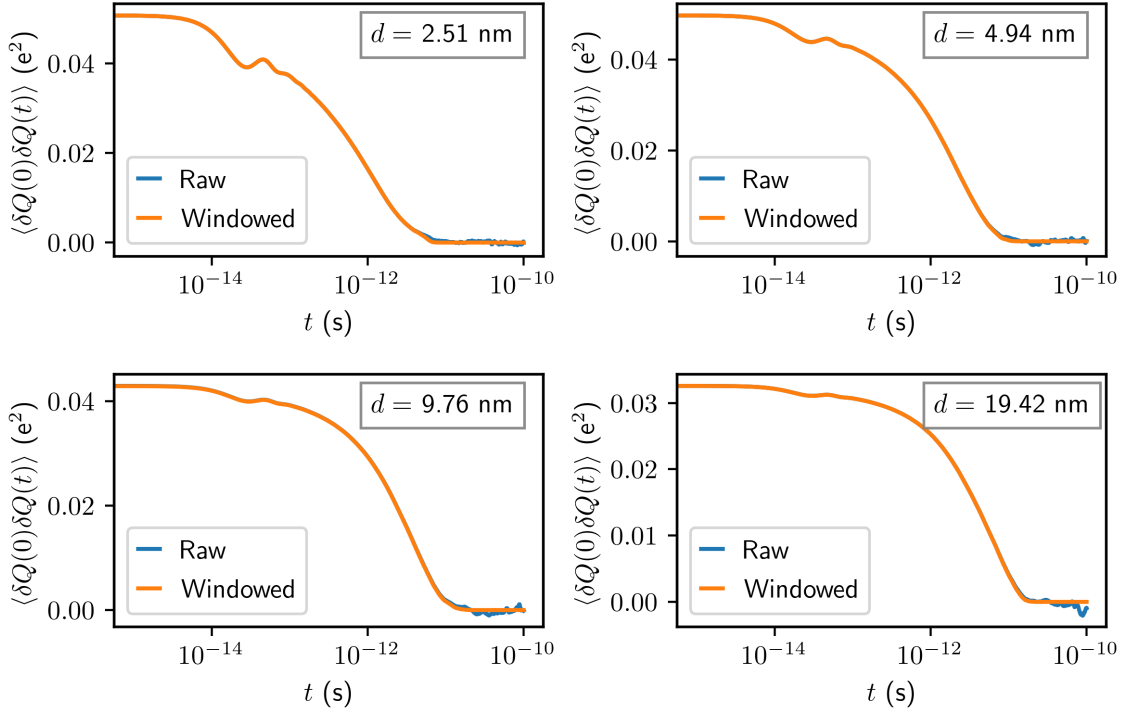

FIG. S5. Comparison between the raw $\langle \delta Q(0) \delta Q(t) \rangle$ as calculated from the MD simulations and the same function after being multiplied by an appropriate window function (see Eq. 8).

could choose a sufficiently small sampling interval until the integrals would converge to a numerically stable solution, but this would require harder computational efforts and the storage of larger amounts of data.

A simple but very effective strategy was introduced by Filon, and it is based on representing the integrand using interpolators, for which the integrals can be calculated analytically [11]. In the specific case of the Fourier-Laplace transform of a signal $f(t)$, we can write the following approximation

$$\hat{f}(\omega) = \int_0^\infty f(t) e^{-i\omega t} dt \approx \sum_{j=0}^{N-1} \int_{t_{kj}}^{t_{kj+k}} v_j(t) e^{-i\omega t} dt, \qquad (9)$$

where the full integral has been substituted with a sum of integrals over consecutive, non-overlapping segments (each one embedding $k+1$ samples of $f(t)$), and the signal has been substituted by a function which interpolates the points sampled from $f(t)$. In our case, we chose to interpolate our points using Lagrange polynomials of order 2 (i.e. in the form of $at^2 + bt + c$) defined on consecutive triplets of data points. With this in mind, Eq. 9 can be

rewritten as a sum over consecutive triplets

$$\hat{f}(\omega) \approx \sum_{j=0}^{N-1} \int_{t_{2j}}^{t_{2j+2}} v_j(t) e^{-i\omega t} dt = \sum_{j=0}^{n-1} \int_{t_{2j}}^{t_{2j+2}} a_j t^2 e^{-i\omega t} dt + \int_{t_{2j}}^{t_{2j+2}} b_j t e^{-i\omega t} dt + \int_{t_{2j}}^{t_{2j+2}} c_j e^{-i\omega t} dt,$$
(10)

where the coefficients $a_j$, $b_j$, $c_j$ are determined from the Lagrange interpolation of $f(t_{2j})$, $f(t_{2j+1})$, $f(t_{2j+2})$. Now the integrals in Eq. 10 can be computed analytically without any numerical error in the quadrature. In this method, we only rely on the assumption that the interpolating functions are sufficiently accurate to represent the original signal. In practice, we first determine the polynomial coefficients by Lagrange interpolation of $\langle \delta Q(0) \delta Q(t) \rangle$ considering consecutive triplets, then use Eq. 10 for the calculation of the Fourier-Laplace transform of $\langle \delta Q(0) \delta Q(t) \rangle$ in the estimation of the admittance. We employ a similar formula for the calculation of the inverse Fourier-Laplace transform, in the analysis of the time response of equivalent circuit models (see next section).

## V. ELECTRIC PROPERTIES

### A. Equivalent circuit model

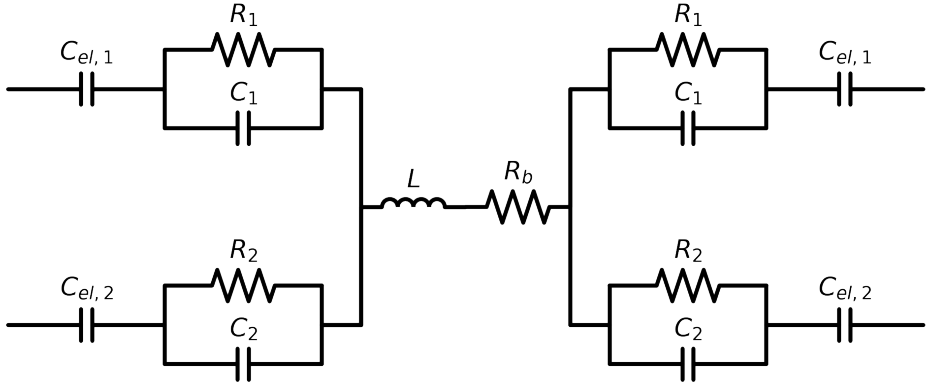

FIG. S6. Representation of the equivalent circuit used to fit the MD results.

All the considered systems display similar features both in terms of charge autocorrelation functions and in terms of the corresponding admittance/impedance. Therefore, we chose to fit the same equivalent circuit (EC) model, depicted in Fig. S6, for all the systems. Despite

the apparent complexity, the choice of our EC model can be supported by a few observations. As can be noted, the structure is almost perfectly symmetrical, thus reflecting the systems' symmetry. The leftmost and rightmost part represent the two interfaces and the respective electrodes. Each interface is represented by a two-branched RC circuit. Each branch roughly represents a single exponential mode in the charge autocorrelation function. Each mode is then 'weighted' by the terminal electrode capacitance $C_{el,1}$. The inductor $L$ was introduced to represent the ripples observed in the charge autocorrelation function at short times, and to fit the high frequency branches in the Nyquist/Bode plots. Finally, the resistor $R_b$ tunes the magnitude of the admittance peak observed in the Bode plots, and serves as a simple shift in the real axis of the Nyquist plot (parametric plot of the impedance in the complex plane, as a function of the frequency). The impedance of this circuit is

$$Z(\omega) = i\omega L + R_b + 2\left[\frac{i\omega C_{el,1} - \omega^2 C_{el,1} R_1 C_1}{i\omega C_{el,1} R_1 + i\omega R_1 C_1 + 1} + \frac{i\omega C_{el,2} - \omega^2 C_{el,2} R_2 C_2}{i\omega C_{el,2} R_2 + i\omega R_2 C_2 + 1}\right]^{-1}. \quad (11)$$

## B. Fitting

We fitted the EC models to admittance and impedance, as calculated from MD results. Our fitting procedure is entirely defined in the frequency domain and takes both the admittance and impedance into account simultaneously. For fitting a complex function, our approach is to minimize the euclidean distance between the model and the MD results in the complex plane. In the complex plane the distance of a point $A$ to a target value $A^*$ is defined as

$$\|\mathbf{A} - \mathbf{A}^*\| = \left[(\Re[A] - \Re[A^*])^2 + (\Im[A] - \Im[A^*])^2\right]^{1/2}, \quad (12)$$

where $\Re$ and $\Im$ are the real and imaginary parts, respectively. In order to fit the MD results, we minimize the following loss function

$$L(\mathbf{x}) = w_Z \frac{\sum_\omega \|Z(\omega,\mathbf{x}) - Z^*(\omega)\|}{\sum_\omega \|Z^*(\omega)\|} + w_Y \frac{\sum_\omega \|Y(\omega,\mathbf{x}) - Y^*(\omega)\|}{\sum_\omega \|Y^*(\omega)\|}, \quad (13)$$

where $\mathbf{x}$ is a vector containing the EC parameters, $Z(\omega,\mathbf{x})$ and $Y(\omega,\mathbf{x})$ are the EC impedance and admittance, respectively, and $w_Z$ and $w_Y$ are weights used to tune the focus on the impedance or on the admittance during the fitting. The weights are manually optimized for each system in order to find a good compromise between the fitting of admittance and impedance. The optimal parameters for the EC models are found as $\mathbf{x}^{opt} = \arg\min L(\mathbf{x})$.

## C. Comparison with simpler EC models

The MD impedance results can be fitted with many possible circuit models. We found our EC to accurately represent all the features present in the impedance/admittance spectra, but one might question what happens if we use a simpler model to fit the MD results. In Fig. S7 we compare different EC models used to fit the impedance results, reported in a Nyquist plot. Note that each EC model is fitted independently, using the loss function introduced in the previous section.

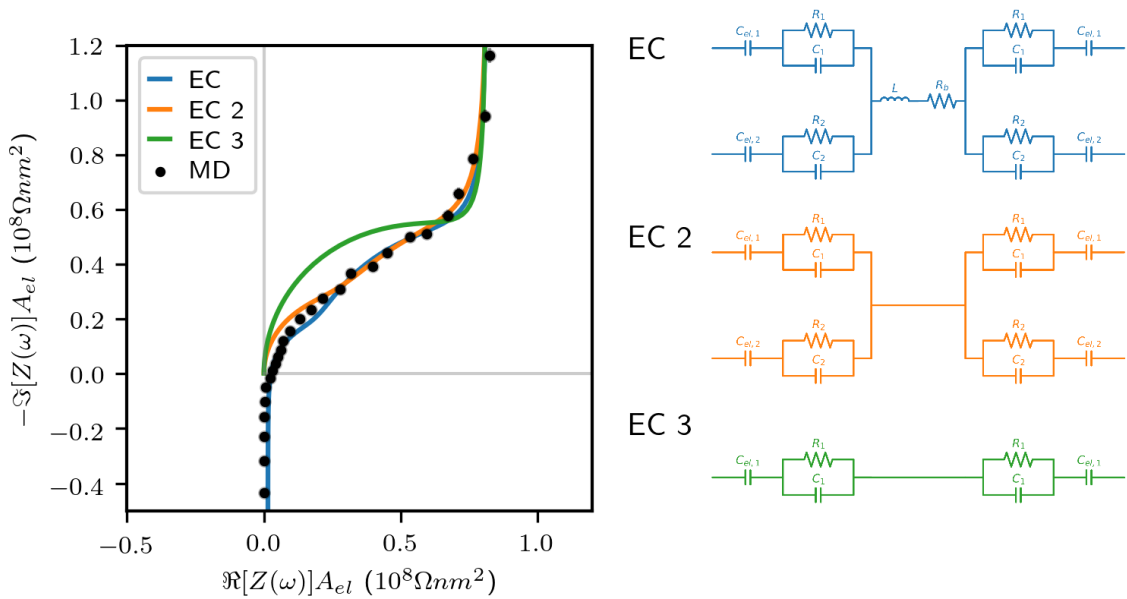

FIG. S7. Comparison of different EC models (right) and their impedance fitted on MD results, represented in a Nyquist plot (left). MD results are reported as black dots. The impedance of each EC model is represented as a solid line. The circuit structures are reported on the right, with the colors matching the ones used in the Nyquist plot.

As expected, simplifying the original model, labelled as 'EC', reduces the accuracy of the impedance fit. Specifically, removing the inductor and resistor ('EC 2') results in the disappearance of the negative branch at high frequencies, which corresponds to neglecting the oscillations present in the total charge autocorrelation function at short times. In 'EC 3', we further remove one of the two branches from each side of the equivalent circuit. In this case, we obtain a single semicircle in the Nyquist plot, which poorly represents the

impedance behaviour at intermediate frequencies.

### D. Electrochemical results

#### 1. Step response in time domain

The equilibrium $\langle \delta Q(0) \delta Q(t) \rangle$ function can be compared with the step response of the EC models. This mainly serves as a validation of the fitted models in the time domain. The charge autocorrelation function is related to the the admittance via

$$\langle \delta Q(0) \delta Q(t) \rangle = \mathcal{L}^{-1} \left\{ \frac{Y(\omega)}{\beta \omega^2} \right\}, \tag{14}$$

where $\mathcal{L}^{-1}$ is the inverse Fourier-Laplace transform. Following the approach used by Straube *et al.* for the definition of the inverse transform [12] this equation can be written in a more explicit form

$$\langle \delta Q(0) \delta Q(t) \rangle = \frac{1}{\pi \beta} \int_{-\infty}^{\infty} \frac{\Re[Y(\omega)]}{\omega^2} e^{-i\omega t} d\omega. \tag{15}$$

We numerically calculated the inverse transform using the Filon-Lagrange method explained in section IV B, swapping the frequency and time domain. The numerical results for all the systems are shown in Fig S8. In all the cases, the equivalent circuit behaviour is in excellent agreement with the MD results.

#### 2. Low frequency resistance and cell capacitance

The total resistance at low frequency can be calculated from the real part of the impedance $R_{\omega \to 0} = \Re[Z(\omega \to 0)]$. In our case, the lowest accessible frequency is $\omega = 1.26 \times 10^{11}$ rad/s. The EC and MD results for this property are shown in Fig. S9a. The equivalent circuit results are in excellent agreement with MD, and both the datasets display a linear trend $R \propto d$, crossing 0 resistance for $d = 0$. This result is in agreement with the macroscopic linear relation between the resistance and the width of a resistor.

From the equivalent circuit models, the capacitance of the cell can be calculated as $C = (C_{el,1} + C_{el,2})/2$. In Fig. S9b we compare the equivalent circuit capacitance with the differential capacitance calculated from the variance of the total charge ($\beta \langle \delta Q^2 \rangle$, from Ref. 13), which corresponds to the capacitance calculated under the Born-Oppenheimer

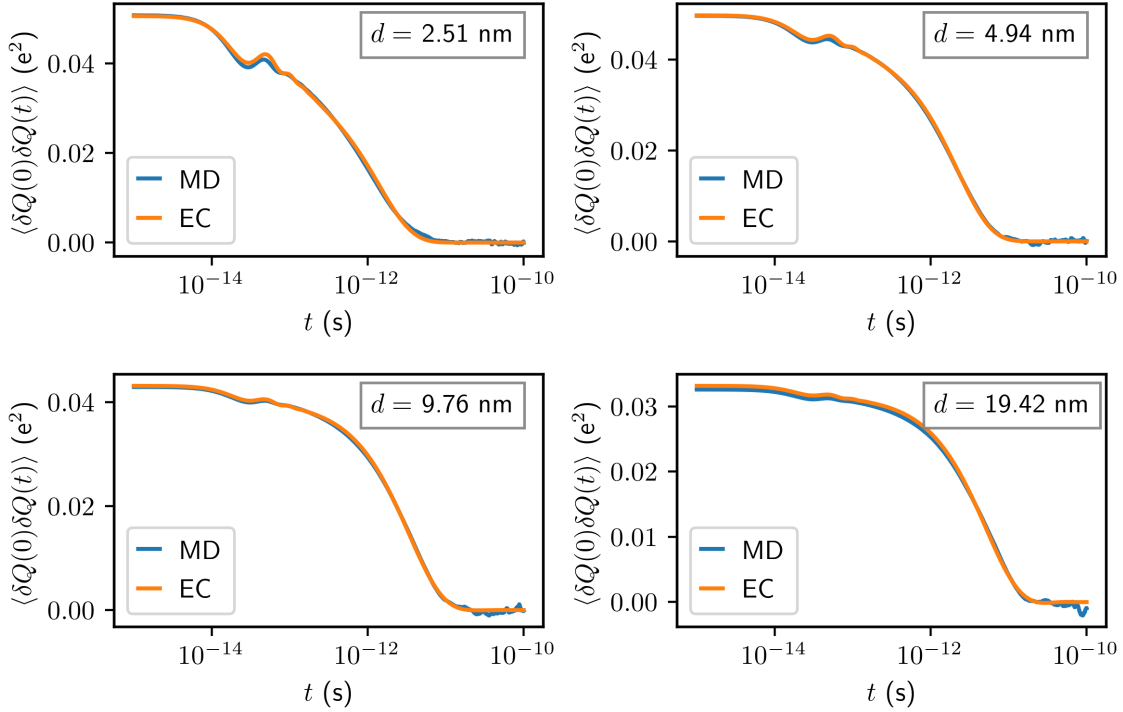

FIG. S8. Comparison between the charge autocorrelation functions as calculated from MD simulations (MD), and as calculated from the equivalent circuit models (EC), for all the systems.

approximation. Overall, we observe a decrease in the cell capacitance as we increase the electrode-electrode separation, as expected for macroscopic capacitors. In all cases, the EC results are in good agreement with the capacitance results from MD simulations.

3. *Charge relaxation time*

Fig. S10 shows the relaxation time $\tau$ of the charge as a function of the inter-electrode distance. $\tau$ can be calculated from the integral of the normalized total charge autocorrelation function. In practice, we perform the integral on the autocorrelation function multiplied by the same window function used for the numerical Laplace transforms to mitigate the effect of the noise at long times. We also estimate $\tau$ as the product between the total resistance (estimated from the impedance as $\Re[Z(\omega \to 0)]$, as calculated from MD results) and the cell capacitance, calculated as $\beta \langle \delta Q^2 \rangle$. The increase of $\tau$ with increasing inter-electrode distance reflects the slower relaxation of $\langle \delta Q(0) \delta Q(t) \rangle$. In all cases, we observe a good agreement between the values of $\tau$ obtained as $RC$ or from the charge ACF.

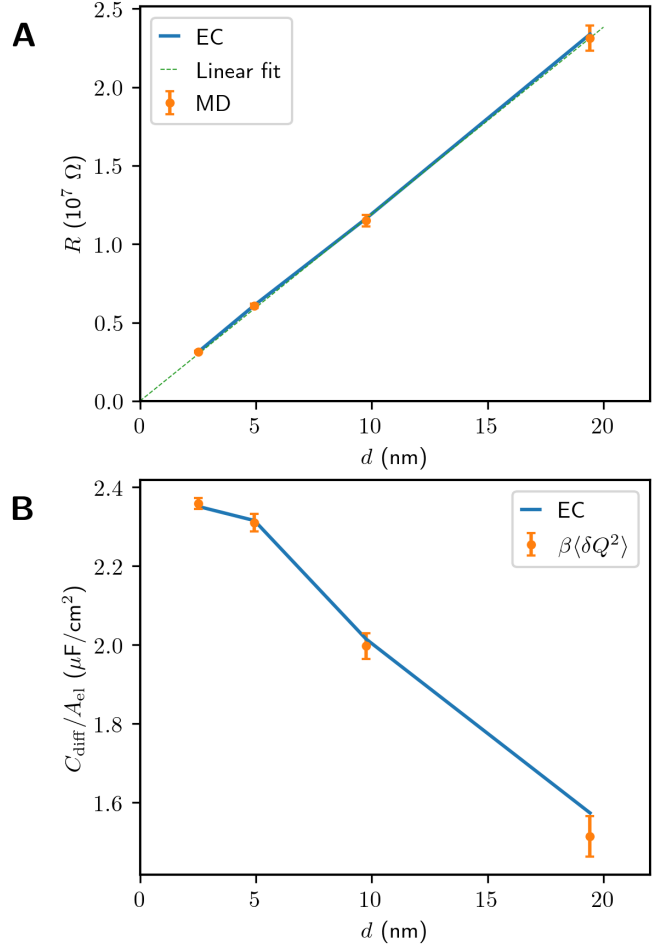

FIG. S9. (a) Total resistance at low frequency as a function of the electrode-electrode distance, as calculated from MD (symbols) and EC models (solid line). The results are compared with a linear fit of the MD results to show the extrapolation at $d = 0$. (b) Comparison between cell capacitance calculated from MD using the electrode charge fluctuations $\beta \langle \delta Q^2 \rangle$ (symbols) and from EC models (solid line) for all the considered inter-electrode distances.

The DDS picture further provides insights into the effect of the inter-electrode distance $d$, thanks to the scalings discussed above or in the main text. Using $R = \alpha d/A_{el}$ (with $\alpha$ a constant) for the cell resistance and the DDS model of the cell capacitance, $A_{el}/C = 2w_{\text{DDS}}/\varepsilon_0 + d_{\text{DDS}}/\varepsilon_0\varepsilon_w$ and $d = d_{\text{DDS}} + 2w_{\text{DDS}}$, we obtain:

$$\tau = RC = \alpha\varepsilon_0\varepsilon_w \frac{d}{d + 2(\varepsilon_w - 1)w_{\text{DDS}}} = \tau_\infty \frac{d}{d + 2(\varepsilon_w - 1)w_{\text{DDS}}} \quad (16)$$

with $\tau_\infty$ the value extrapolated for $d \to \infty$. Fig. S10 shows that this scaling rather accurately describes the evolution of the relaxation time with $d$. Fitting the results with Eq. 16, we

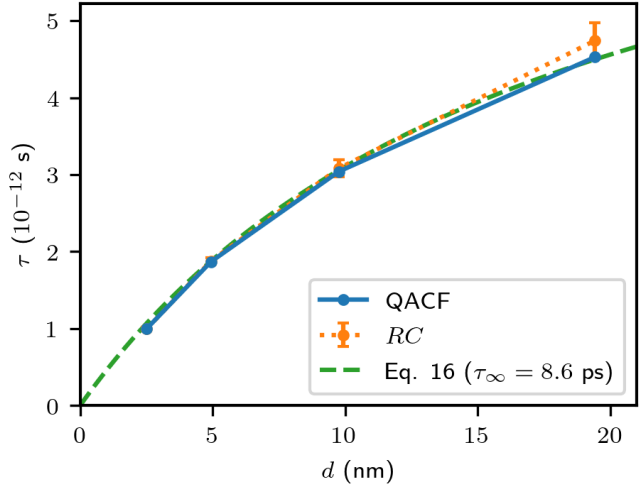

FIG. S10. Charge relaxation time as a function of the inter-electrode distance. $\tau$ is estimated either as the integral of the normalized total charge ACF (blue solid line) or as the $RC$ time (orange dotted line), using the total resistance at low frequency and the cell capacitance. The scaling of Eq. 16 is also shown (dashed green line).

obtain $\tau_\infty \approx 8.6$ ps. This value is close to the characteristic time corresponding to the Debye relaxation in bulk water ($\tau_D \approx 9$ ps) that can be interpreted as arising from the migration of orientational defects in the H-bond network [14].

### 4. Individual parameters

The trends of the equivalent circuit parameters are reported in Fig. S11. In general, we note that all the capacitances tend to decrease as we increase the electrode-electrode separation, as it is expected from macroscopic theory. In contrast, the resistances tend to increase. This is also expected, since wider dielectric slabs are associated with larger values of resistances. We finally note a clear linear scaling the inductance as a function of $d$.

## VI. TOTAL DIPOLE MOMENT AND ELECTRODE CHARGE

For a given configuration, the total dipole moment $\mathbf{M}$ of the system is

$$\mathbf{M}_{\text{tot}}(t) = \sum_{i \in \text{water}} q_i \mathbf{r}_i(t) + \sum_{j \in \text{electrode}} q_j(t) \mathbf{r}_j \;, \qquad (17)$$

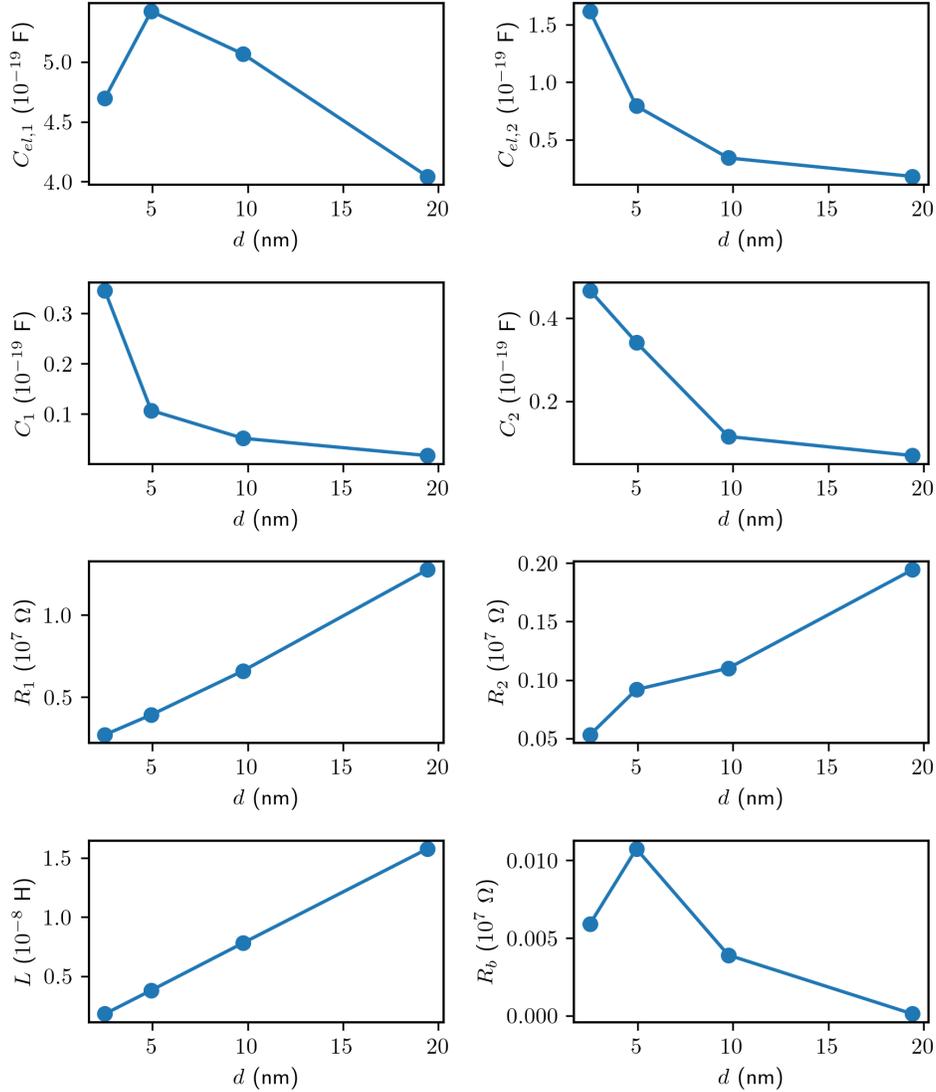

FIG. S11. Equivalent circuit parameters for the four systems we considered.

where the first sum runs over the mobile atoms of the liquid with fixed partial charges, while the second runs over the fixed electrode atoms with fluctuating charges, with $q$ and $\mathbf{r}$ denoting the atomic charges and positions, respectively. Note that for the calculation of $\mathbf{M}$, we consider the unwrapped coordinates, to avoid artefacts due to periodic boundary conditions. From the symmetry of the system, one should consider the components of the dipole moment parallel ($\parallel$) and perpendicular ($\perp$) components to the electrode surfaces. With our choice of axes, the latter simply corresponds to the $z$ component, and we estimate the autocorrelation function (ACF) of the parallel component as the average of the ACF of the $x$ and $y$ components.

Separating the perpendicular component into the water and electrode dipoles, and separating the latter into the contributions of the two electrodes, we have:

$$M_{\text{tot}}^{\perp} = M_{\text{wat}}^{\perp} + M_{\text{el}}^{\perp} = M_{\text{wat}}^{\perp} + Q_1(t)z_{Q,1}(t) + Q_2(t)z_{Q,2}(t) \tag{18}$$

where in the second line we introduced the total charge of each electrode, as well as the charge-weighted positions of each electrode (which are well defined except when the electrode charge vanishes). Due to the global electroneutrality of the system, at each time step $Q_1(t) = -Q_2(t) = Q(t)$, so that $M_{\text{tot}}^{\perp} = M_{\text{wat}}^{\perp} - Q(t)d_Q(t)$, where we introduced the difference $d_Q(t) = z_{Q,2}(t) - z_{Q,1}(t)$. For a potential difference between the electrodes of 0 V (shortcut conditions), the total dipole in the direction perpendicular to the electrodes vanishes, and we obtain, for each configuration,

$$M_{\text{wat}}^{\perp}(t) = Q(t)d_Q(t) \ . \tag{19}$$

Since the electrodes are treated as perfect metal (for the description of screening within the metal with fluctuating charges in classical MD simulations, see Ref. 15), we expect the charge to be localized mainly in the first atomic plane in contact with the liquid, regardless of the charge distribution within that first plane. Therefore, one can approximate $d_Q(t)$ by the time-independent distance $d$ between the first atomic planes of both electrodes. Fig. S12 demonstrates that the resulting proportionality between $\delta M_z = M_{wat}^{\perp} - \langle M_{\text{wat}}^{\perp}\rangle = M_z$ and $Q(t)$ is very well satisfied and that the effective distance $d_{\text{eff}}$ obtained from a linear fit of the correlation corresponds to within less than 1% to the the actual distance $d$ between the first atomic planes of both electrodes. This figure shows in contrast no correlation between the total charge of the electrodes and the parallel components of the total water dipole, as expected.

Following Eq. 19 and the observation that $d_Q(t) \approx d$, one expects that the perpendicular component of the total water dipole moment follows the same dynamics as the total electrode charges. Fig. S13 shows an almost perfect match between the ACFs of $Q$ and $M_{\text{wat}}^{\perp}$, while $\langle \delta M_{\text{wat}}^{\|}(0)\delta M_{\text{wat}}^{\|}(t)\rangle$ decays more slowly and only marginally depends on $d$, with a slight decrease in the relaxation time with increasing $d$, consistently with the results reported in Ref 16.

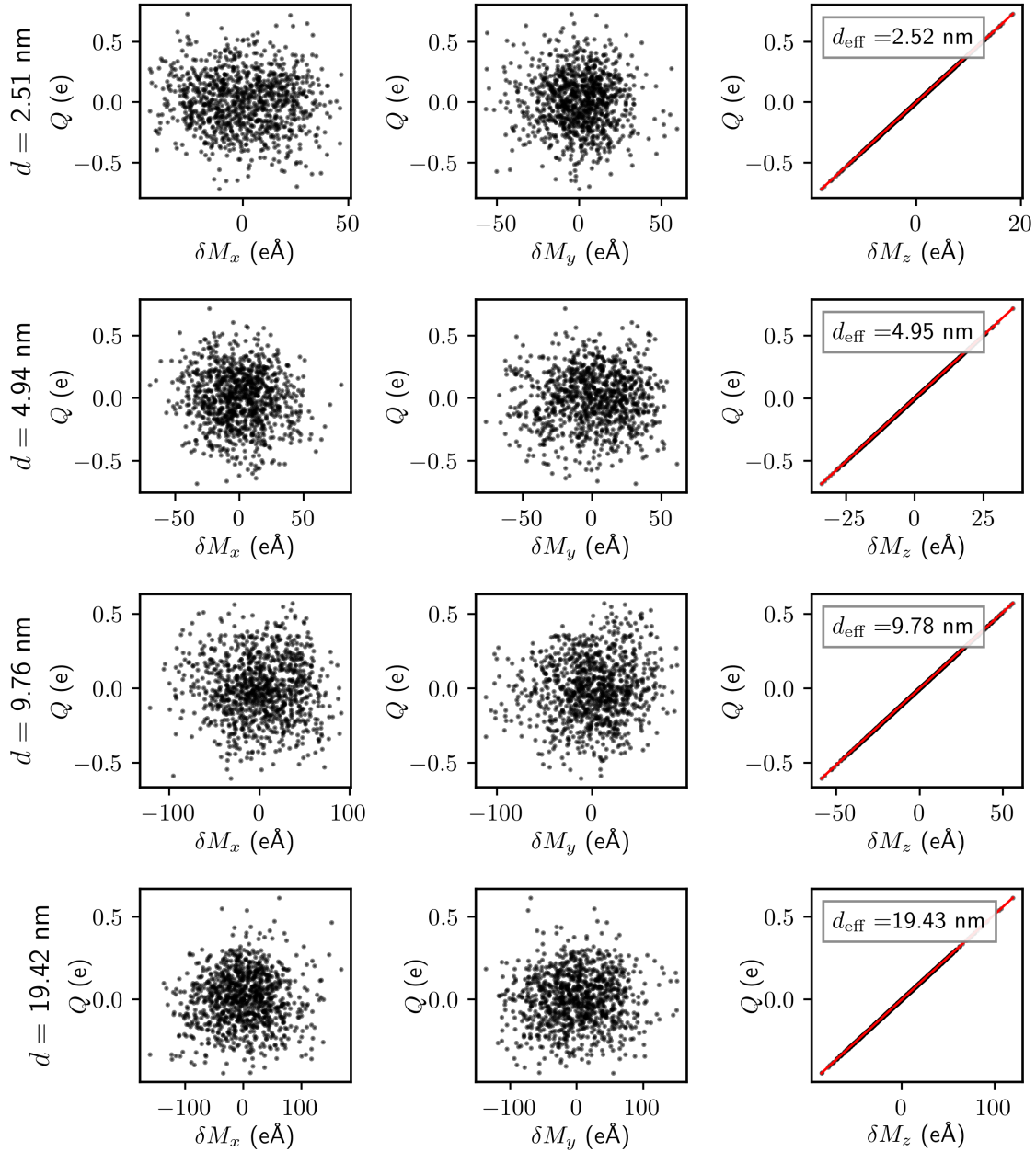

FIG. S12. Correlation plots between the total charge and the components of total dipole moment of water, for all the considered systems. The total dipole moment values are reported as fluctuations (instantaneous value subtracted by the ensemble average value). Each row represents a system, identified by the electrode-electrode distance (reported on the left). The three columns report the $x$, $y$ and $z$ component of $\mathbf{M}_{\text{wat}}$, respectively. In the third column we also report the effective electrode-electrode distance found from the slope of a line fitted on the data.

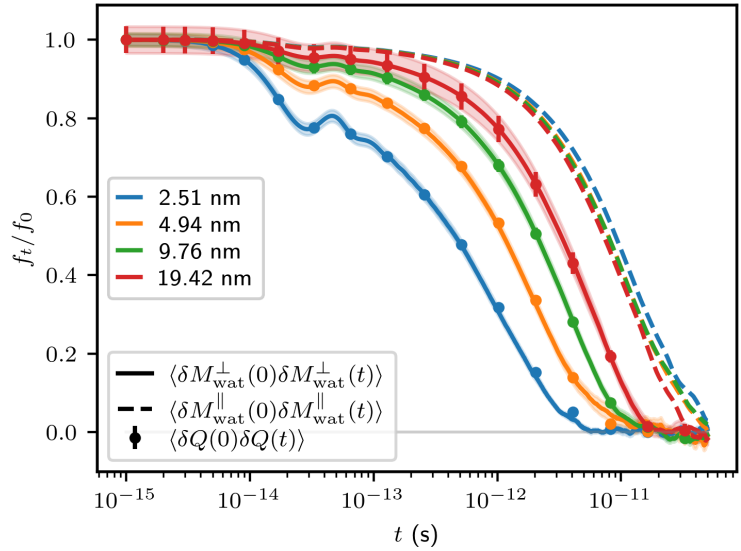

FIG. S13. Autocorrelation functions of the dipole moment of the confined water slab, with components $M_{\text{wat}}^{\perp}$ (solid lines) and $M_{\text{wat}}^{\parallel}$ (dashed lines) perpendicular and parallel to the electrode surfaces, and of the electrode charge $Q$ (symbols), for the 4 considered inter-electrode distances (colors). The uncertainty, quantified as the standard error from 10 blocks of each trajectory, is shown as shaded areas for $M_{\text{wat}}^{\perp}$ and as error bars for $Q$; it is omitted for $M_{\text{wat}}^{\parallel}$ for clarity.


}

\end{document}